\documentclass{ifacconf}

\usepackage{graphicx}
\usepackage{natbib}
\usepackage{amsmath,amssymb}
\usepackage{mathtools}
\usepackage{xcolor}
\usepackage[normalem]{ulem} %
\usepackage{algorithm}
\usepackage[noend]{algpseudocode}

\newcommand{\E}{\mathbb{E}}
\renewcommand{\qed}{\leavevmode\unskip\penalty9999 \hbox{}\nobreak\hfill\quad\hbox{$\Box$}} %
\DeclareMathOperator*{\argmax}{arg\,max}

\newtheorem{problem}{Problem}

\begin{document}
\begin{frontmatter}

\title{Universal Quantizer Design for Identification of Systems with Memory\thanksref{footnoteinfo}}

\thanks[footnoteinfo]{This work has been partially supported by the Swedish Research Council under contract number 2023-05170 and by the Wallenberg AI, Autonomous Systems and Software Program (WASP). } %

\author[KTH]{Yibo Shi}
\author[KTH]{Ying Wang}
\author[KTH]{Karl H. Johansson}
\author[KTH]{Cristian R. Rojas}

\address[KTH]{Department of Decision and Control Systems, KTH Royal Institute of Technology, SE-100 44 Stockholm, Sweden (e-mail: yibos@kth.se, yinwa@kth.se, kallej@kth.se, crro@kth.se).}

\begin{abstract} %
In system identification from quantized observations, the quantizer determines how much information about the parameters the observations retain, and hence the achievable estimation accuracy. %
Quantizer design can be guided by the Fisher information matrix (FIM), which assesses the parameter information retained by candidate quantizers. Existing methods, however, typically rely on closed-form expressions of the FIM.
This paper proposes a universal quantizer design method for general systems with memory. 
Its core is a universal FIM estimator that evaluates the information per bit from simulated trajectories, and the thresholds are designed by maximizing the estimated criterion under a bit budget. 
For unknown parameters, the design is implemented through a two-stage procedure.
Numerical studies on a quantized ARX system yield designs close to the reference optimum, while a nonlinear cascaded-tanks example reduces the identification error by about 50\% relative to a single median threshold.
\end{abstract}


\end{frontmatter}

\section{Introduction}
\label{sec:introduction}

System identification builds models of dynamical systems from measured input and output signals, and it usually assumes that the signals are measured with high precision~\citep{ljung1999system}. In many cases, however, only a quantized output is available. A quantizer compares the output with a few thresholds and reports only the interval that contains it. Quantized outputs arise from low-cost sensors, such as the binary sensors of industrial plants and automotive systems~\citep{wang2003system, casini2011input}, and from communication channels with limited bandwidth in sensor networks and networked control systems~\citep{wang2008space, ribeiro2006bandwidth, kar2012optimal}. Identification from quantized observations has therefore been studied extensively, and consistent and asymptotically efficient estimators are available for binary and multilevel sensors~\citep{wang2003system, wang2010system,wang2024asymptotically, godoy2011identification}. The accuracy of any unbiased estimator is limited by the \emph{Cram\'er--Rao lower bound} (CRLB), which is the inverse of the \emph{Fisher information matrix} (FIM) of quantized observations. For a given system, the FIM further depends on the experimental conditions, that is, the input signal and the quantizer. Optimal experiment design~\citep{goodwin1977dynamic, pukelsheim2006optimal} chooses these conditions so as to maximize a scalar criterion of the FIM. For quantized identification, it has mainly addressed the input signal~\citep{casini2011input, casini2012input, godoy2014novel}, whereas the design of the quantizer, called \emph{quantizer design} in this paper, has received far less attention.

Quantizer design consists of two problems, which are \emph{threshold selection} and \emph{resource allocation}~\citep{wang2008space}. Threshold selection places a given number of thresholds so as to maximize the design criterion. A binary sensor produces one bit per observation, and a quantizer with more thresholds produces more bits. Under a budget of bits, such as the one set by a limited bandwidth, a finer quantizer therefore leaves fewer observations. Resource allocation chooses the number of thresholds and the number of observations under this budget.

For FIR and OE systems, the FIM has a closed-form expression in terms of the thresholds~\citep{wang2007asymptotically, marelli2013identification}, and both problems have been solved for FIR systems~\citep{wang2008space, wang2024threshold}. In general, no such expression exists for \emph{systems with memory}, namely systems in which noise enters the dynamics, such as ARX and ARMAX models and state-space models with process noise, because their likelihood is a marginal over a hidden state~\citep{kantas2015particle}. Parameter estimators are available for ARX systems with quantized outputs~\citep{csaji2012recursive, aguero2017em}. To the best of our knowledge, however, quantizer design has not been solved for systems with memory, and \citet{wang2024threshold} name this extension as future work.

For a system with memory, quantizer design therefore requires the FIM to be computed numerically for each candidate quantizer. The available numerical methods are restricted to particular classes of models. Point-mass filters, which discretize the state on a grid~\citep{bucy1971digital}, require a state of low dimension, and particle estimates of the score require the transition density of the state~\citep{poyiadjis2011particle, valenzuela2017robust}. The proposed method estimates the FIM from simulated trajectories alone, which any model can provide. It uses the universal FIM estimator of~\citet{shi2026universal}, which is based on \emph{context-tree weighting} (CTW) and needs no model-specific expression. That estimator was developed for sources without inputs and with finite memory, and it is extended in this paper to systems with inputs and a hidden state. The thresholds of the quantizer are then chosen to maximize the estimated criterion. As a function of the thresholds, this criterion is piecewise constant, because the quantized data changes only when a threshold crosses a simulated output. Gradient-based methods are therefore not directly applicable, and the criterion is maximized by a direct search~\citep{kolda2003optimization}.
A bound relates the loss of the resulting design to the error of the FIM estimate. Finally, the best quantizer depends on the true parameter, as in any locally optimal experiment design~\citep{goodwin1977dynamic}. A design for the worst case over a prior set~\citep{wang2008space, wang2024threshold} would require the criterion at many parameter values, each at the cost of extra simulations. A two-stage design~\citep{pronzato2013design} is used instead, which first estimates the parameter with part of the bit budget.

The contributions of this paper are as follows:
\begin{itemize}
\item A universal quantizer design method for systems with memory that needs only simulated trajectories, with a bound on the loss of the designed quantizer.
\item A two-stage design under a bit budget for an unknown parameter.
\item A numerical validation on a quantized ARX system and on a model of cascaded tanks.
\end{itemize}

The problem is formulated in Section~\ref{sec:problem}, and the design method is presented in Section~\ref{sec:method}. The numerical study is reported in Section~\ref{sec:numerics}, and conclusions are drawn in Section~\ref{sec:conclusion}.

\section{Problem Formulation}
\label{sec:problem}

\subsection{System and sensor} The system has an unknown parameter $\theta \in \Theta \subset \mathbb R^d$, where $\Theta$ is compact, with true value $\theta_0 {} \in \operatorname{int} \Theta$, and is described by
\begin{equation}
x_t = a_\theta(x_{t-1}, u_t, w_t), \qquad y_t = c_\theta(x_t) + v_t ,
\label{eq:system}
\end{equation}
where $u_t$ is the input, $x_t$ is the state, and $y_t$ is the output. The noise $(w_t, v_t)$ is i.i.d.\ and independent of the input. The input is a given stationary ergodic process on a finite set $\mathcal U$, for instance an i.i.d.\ multilevel signal, and its law does not depend on $\theta$. 
For a given parameter value $\theta$, a simulator of~\eqref{eq:system} can generate trajectories $(u_t, y_t)$.

During identification, the output $y_t$ is observed only through a sensor, namely a quantizer with thresholds $C = (C_1, \dots, C_m)$, where $C_1 < \dots < C_m$. The number $m$ of thresholds sets the resolution of the sensor. The thresholds split the real line into the $m + 1$ intervals $(-\infty, C_1], (C_1, C_2], \dots, (C_m, \infty)$, called \emph{cells}. The sensor reports the index of the cell that contains $y_t$, namely $q_t = \mathcal Q_C(y_t) := {\lvert \{ i : C_i < y_t \} \rvert}$~\citep{wang2008space, wang2024threshold}. The thresholds range over the compact set
\begin{equation}
\mathcal C_m := \bigl\{ C \in [c_{\min}, c_{\max}]^m :\ C_{i+1} - C_i \ge \nu,\ i < m \bigr\},
\label{eq:feasible}
\end{equation}
with range $[c_{\min}, c_{\max}]$ and minimum separation $\nu > 0$ between adjacent thresholds. The number $m$ of thresholds is at most $M$, with $(M - 1)\nu \le c_{\max} - c_{\min}$, so that every $\mathcal C_m$ is nonempty. The observations are $z_t = (u_t, q_t)$, which contain the output only through the index $q_t$. Identification uses $N$ observations $z_1, \dots, z_N$, and $N$ is limited by the bits that the sensor reports.

\subsection{Design criterion} For fixed $C$, let $I^{(N)}(\theta; C)$ be the FIM of the observations $z_{1:N}$, and let $I(\theta; C) := \lim_{N} I^{(N)}(\theta; C)/N$ be the \emph{information rate}. This limit exists under the following conditions, which are assumed for $\theta$ near $\theta_0$ and $C \in \mathcal C_m$.
\begin{itemize}\itemsep0pt
\item[(A1)] The process $(u_t, y_t)$ is stationary and ergodic, and the conditional law $P_\theta(q_t \mid u_t, z_{t-1}, z_{t-2}, \dots)$ of $q_t$ given $u_t$ and the past observations is smooth in $\theta$.
\item[(A2)] For $L \ge 1$, let the \emph{score} of the conditional law of $q_t$ be $\ell_t^L := \nabla_\theta \log P_\theta(q_t \mid u_t, z_{t-1}, \dots, z_{t-L+1})$. These scores forget the distant past at a geometric rate $\rho \in (0, 1)$, namely $\sup_{L' \ge L} (\E \lVert \ell_t^{L'} - \ell_t^{L} \rVert^2)^{1/2} = O(\rho^L)$.
\item[(A3)] Every input level and every cell have probability bounded away from zero, uniformly over the past.
\end{itemize}
Forgetting as in (A2) holds for hidden Markov models under mixing conditions~\citep{legland2000exponential, douc2011consistency}. Assumption (A3) requires the separation $\nu$ in~\eqref{eq:feasible}, because coincident thresholds leave a cell of zero probability. It also requires $[c_{\min}, c_{\max}]$ to lie inside the range of $y_t$, as in~\citet{wang2008space}.

For $N$ observations, the CRLB is approximately equal to $I(\theta; C)^{-1}/N$. Under regularity conditions, the maximum likelihood estimator attains this bound asymptotically~\citep{bickel1998asymptotic, douc2004asymptotic}. Unlike the information $I^{(N)}(\theta; C)$ of $N$ observations, the information rate $I(\theta; C)$ depends neither on $N$ nor on the initial state, so the design is based on the rate. A design criterion $\Psi$~\citep{pukelsheim2006optimal} maps the information matrix to a scalar to be maximized. By default we use E-optimality, $\Psi(I) = \lambda_{\min}(I)$, which matches the spectral-norm index of~\citet{wang2024threshold}. With $m + 1$ cells, the sensor reports each index $q_t$ with $\log_2(m+1)$ bits~\citep{wang2008space}. A \emph{bit budget} $R$ limits the total number of bits that the sensor reports during identification, so the number $N$ of observations satisfies $N \log_2(m+1) \le R$.

\subsection{Problem statement}
\begin{problem}[Quantizer design]
\label{prob:design}
The design problem is
\begin{equation}
\max_{m \le M}\ \max_{C \in \mathcal C_m}\ \max_{N \le R/\log_2(m+1)}\ \Psi\bigl(N\, I(\theta_0; C)\bigr),
\label{eq:allocation}
\end{equation}
that is, resource allocation, which chooses the number $m$ of thresholds, the thresholds $C$, and the number $N$ of observations under the bit budget.
\end{problem}

Because $\Psi$ is nondecreasing in the Loewner order and positively homogeneous~\citep{pukelsheim2006optimal}, the maximization over $N$ in~\eqref{eq:allocation} uses up the budget, i.e., $N = R/\log_2(m+1)$ up to rounding, and its value is $R\, J(\theta_0; C)$, where
\begin{equation}
J(\theta; C) := \Psi\bigl( I(\theta; C) / \log_2(m+1) \bigr)
\label{eq:perbit}
\end{equation}
is the \emph{criterion per bit}. The resource allocation problem~\eqref{eq:allocation} is thus equivalent to maximizing $J(\theta_0; C)$ over $m \le M$ and $C \in \mathcal C_m$, and its maximizers do not depend on $R$. For a given $m$, the factor $\log_2(m+1)$ only rescales $\Psi\bigl(I(\theta_0; C)\bigr)$, so the maximization over $C$ in~\eqref{eq:allocation} is equivalent to threshold selection for a sensor with $m$ thresholds,
\begin{equation}
\max_{C \in \mathcal C_m}\ \Psi\bigl(I(\theta_0; C)\bigr),
\label{eq:selection}
\end{equation}
whose maximizers are denoted by $C_m^\star$. Resource allocation therefore solves~\eqref{eq:selection} for each $m \le M$ and compares $J(\theta_0; C_m^\star)$ across $m$.

Solving~\eqref{eq:selection} requires $I(\theta_0; C)$ for every candidate $C$. With noise in the dynamics of~\eqref{eq:system}, $I(\theta; C)$ generally has no closed-form expression, and it depends on the unknown $\theta_0$, as in locally optimal design~\citep{goodwin1977dynamic}.

\section{Quantizer Design}
\label{sec:method}

In Section~\ref{sec:estimate}, we estimate the information rate $I(\theta; C)$ at a given parameter value $\theta$ from simulated trajectories. We then maximize the estimated criterion over the thresholds in Section~\ref{sec:search} and obtain this parameter value from the plant by a two-stage design in Section~\ref{sec:twostage}.

\subsection{Estimating the criterion}
\label{sec:estimate}

The information rate $I(\theta; C)$ is the curvature at $\theta$ of the Kullback--Leibler (KL) divergence rate $D(\theta \Vert \theta')$ between the laws of $z$ under $\theta$ and $\theta'$. The estimator, adapted from~\citet{shi2026universal}, simulates trajectories of~\eqref{eq:system} at $\theta$ and at nearby probe points, and it estimates the KL rates between them with context-tree weighting (CTW) models~\citep{willems1995context}.

Under (A1)--(A3), for a step size $h > 0$ and a unit vector $\xi \in \mathbb R^d$, $D(\theta \Vert \theta + h\xi) = \tfrac12 h^2 \xi^\top I(\theta; C)\, \xi + h^3 \tau(\xi) + O(h^4)$, where $\tau$ is a cubic form in $\xi$~\citep{shi2026universal}. Since $\tau(-\xi) = -\tau(\xi)$, the expansions at $\pm h\xi$ add up to
\begin{equation}
D(\theta \Vert \theta + h\xi) + D(\theta \Vert \theta - h\xi) = h^2\, \xi^\top I(\theta; C)\, \xi + O(h^4).
\label{eq:curvature}
\end{equation}
The symmetric matrix $I(\theta; C)$ has $r = d(d+1)/2$ distinct entries, so the method uses $r$ directions $\xi_j$ whose forms $\xi_j^\top I \xi_j$ determine it. The points $\theta \pm h\xi_j$ are called \emph{probe points}.

The KL rates are those of the observations $z$, so they are estimated from simulated trajectories that are quantized with the thresholds $C$. Trajectories of length $n$ are generated once and independently: two trajectories $Y$ and $\tilde Y$ at $\theta$, and one trajectory $Y_j^{\pm}$ at each probe point. The parameter value $\theta$ is restricted to the set $\Theta_h := \{\theta' \in \Theta : \theta' \pm h\xi_j \in \Theta,\ j = 1, \dots, r\}$, so that the probe points lie in $\Theta$. For thresholds $C$, quantizing every output gives the \emph{quantized trajectories} $Z = \mathcal Q_C(Y)$, $\tilde Z = \mathcal Q_C(\tilde Y)$, and $Z_j^{\pm} = \mathcal Q_C(Y_j^{\pm})$, which consist of symbols $z_t = (u_t, q_t)$, like the observations. The trajectories do not depend on $C$, so another $C$ only requires quantizing them again.

The KL rate $D(\theta \Vert \theta')$ is identical to the difference between the cross-entropy rate of the law of $z$ under $\theta$ with respect to the law under $\theta'$, and the entropy rate under $\theta$. Both rates are estimated with CTW models. A CTW model of depth $L$ is trained on a quantized trajectory $A$ and then kept fixed. After training, it assigns a probability $\hat T_A(q \mid s)$ to every output value $q$ given a \emph{context} $s$. The context at time $t$ is $s_t = (u_t, z_{t-1}, \dots, z_{t-L+1})$, namely the current input and the $L-1$ most recent symbols. Only the output is predicted, because the input law does not depend on $\theta$ and therefore does not contribute to the KL rate. For another quantized trajectory $B$, the cross-entropy rate of the law of $B$ with respect to the law of $A$ is estimated by $\hat H(A \to B) = -n^{-1} \sum_t \log \hat T_A(q_t \mid s_t)$, where $q_t$ and $s_t$ are taken from $B$.

With $A = Z_j^{\pm}$ and $B = Z$, this estimate gives the cross-entropy rate. With $A = \tilde Z$, which has the same law as $Z$, it gives the entropy rate. The KL rate is therefore estimated by
\begin{equation}
\hat D(\theta \Vert \theta \pm h\xi_j) = \hat H(Z_j^{\pm} \to Z) - \hat H(\tilde Z \to Z),
\label{eq:kl-estimate}
\end{equation}
which is a variant of the estimator of~\citet{cai2006universal}. 

Inserting~\eqref{eq:kl-estimate} into~\eqref{eq:curvature} and dividing by $h^2$ gives an estimate of $\xi_j^\top I(\theta; C)\, \xi_j$ for each $j$, denoted as $v_j$. The matrix $I(\theta; C)$ is then recovered as the least-squares solution of the linear equations $v_j = \xi_j^\top I \xi_j$, $j = 1, \dots, r$. This solution is projected onto the cone of positive semidefinite matrices, which gives the estimate $\hat I_n(\theta; C)$. The estimated criterion is
\begin{equation}
\hat J_n(\theta; C) = \Psi\bigl(\hat I_n(\theta; C) / \log_2(m+1)\bigr).
\label{eq:jhat}
\end{equation}
The estimate $\hat I_n(\theta; C)$ does not depend on the choice of $\Psi$, so it can also be used to evaluate other design criteria.

The process $z$ has infinite memory, because the state of~\eqref{eq:system} is hidden, so no finite context contains the whole past. A model of depth $L$ therefore estimates the information rate $I_L(\theta; C)$ of the depth-$L$ conditional law $P_\theta(q_t\mid s_t)$. Let $J_{h,L}(\theta; C)$ denote the criterion~\eqref{eq:jhat} computed from the exact KL rates of the depth-$L$ conditional laws, called the \emph{limit criterion}. Under (A1) and (A3), for each $C$,
\begin{equation}
\hat J_n(\theta; C) \to J_{h,L}(\theta; C) \quad \text{almost surely as } n \to \infty,
\label{eq:limit}
\end{equation}
because a model of depth $L$ has finitely many contexts, and their frequencies and predictions converge by the ergodic theorem. The limit criterion differs from $J(\theta; C)$ by a bias of order $h^2 + \rho^L$. The term $h^2$ is the remainder in~\eqref{eq:curvature}. The term $\rho^L$ follows from (A2), because $I_L(\theta; C)$ and $I(\theta; C)$ are the second moments of the scores of $q_t$ given the context $s_t$ and given the whole past, respectively.

\subsection{Maximizing the estimated criterion}
\label{sec:search}

For a parameter value $\theta$, replacing $I(\theta_0; C)$ by $\hat I_n(\theta; C)$ in~\eqref{eq:allocation} and~\eqref{eq:selection} gives a problem that can be solved numerically. The trajectories $Y, \tilde Y, Y_j^{\pm}$ do not depend on $C$, so $\hat J_n(\theta; \cdot)$ is a deterministic function of $C$ on $\mathcal C_m$. Maximizing it over $\mathcal C_m$ is therefore a sample-path optimization problem~\citep{robinson1996analysis, shapiro2021lectures}. Resource allocation only requires this maximization for each $m \le M$, as shown after Problem~\ref{prob:design}. 
To account for numerical optimization error, consider a design $\hat C_{m,n} \in \mathcal C_m$ whose estimated criterion is within the tolerance $\varepsilon_{\rm opt} \ge 0$ of the global maximum:
\begin{equation}
\hat J_n(\theta; \hat C_{m,n}) \ge \sup_{C \in \mathcal C_m} \hat J_n(\theta; C) - \varepsilon_{\rm opt}.
\label{eq:design-rule}
\end{equation}
The following proposition relates its loss under the true criterion to the optimization error, the uniform estimation error, and the bias difference between designs.
\begin{prop}[Loss bound]
\label{prop:loss}
Let $\beta(\theta; C) := J_{h,L}(\theta; C) - J(\theta; C)$ be the bias of the limit criterion~\eqref{eq:limit}, and let $e_n := \sup_{C \in \mathcal C_m} \lvert \hat J_n(\theta; C) - J_{h,L}(\theta; C) \rvert$ be the uniform error of the estimate. For any design $\hat C_{m,n}$ that satisfies~\eqref{eq:design-rule} and every maximizer $C_m^\star$ of $J(\theta; \cdot)$ over $\mathcal C_m$,
\begin{equation}
\begin{aligned}
J(\theta; C_m^\star) - J(\theta; \hat C_{m,n}) \le{}& \varepsilon_{\rm opt} + 2e_n\\
&+ \beta(\theta; \hat C_{m,n}) - \beta(\theta; C_m^\star).
\end{aligned}
\label{eq:loss-bound}
\end{equation}
\end{prop}

\begin{pf}
By the definition of $e_n$ and by~\eqref{eq:design-rule}, $J_{h,L}(\theta; C_m^\star) - e_n \le \hat J_n(\theta; C_m^\star) \le \hat J_n(\theta; \hat C_{m,n}) + \varepsilon_{\rm opt} \le J_{h,L}(\theta; \hat C_{m,n}) + e_n + \varepsilon_{\rm opt}$. Substituting $J_{h,L} = J + \beta$ gives~\eqref{eq:loss-bound}.\qed
\end{pf}

The tolerance $\varepsilon_{\rm opt}$ is the accuracy of the search, which the user trades against its cost. The uniform error $e_n$ comes from the finite trajectory length, mainly through the variance of the estimate, and it decreases as $n$ grows. The bias $\beta$ is of order $h^2 + \rho^L$ (Section~\ref{sec:estimate}), and a longer trajectory does not reduce it. However, only the difference $\beta(\theta; \hat C_{m,n}) - \beta(\theta; C_m^\star)$ of the bias between the selected design and the optimal design enters~\eqref{eq:loss-bound}. If the bias is nearly the same for all designs, this difference is small, and the selected design is nearly optimal even when the bias itself is large. The ranking of designs can therefore be more accurate than the values of the estimate.

The estimate $\hat J_n(\theta; \cdot)$ is piecewise constant, because the quantized trajectories change only when a threshold crosses one of the simulated outputs. Its gradient therefore vanishes almost everywhere, and the estimate may have several local maxima. A design that satisfies~\eqref{eq:design-rule} is thus sought by a search that uses only values of $\hat J_n$ and covers $\mathcal C_m$. The search has two parts, namely a \emph{coarse search} on a grid and a \emph{local search} that refines the best grid points.

The coarse search evaluates the estimate on a grid in $\mathcal C_m$ with few points per coordinate. To reduce its cost, it uses the estimate $\hat J_{n_c}(\theta; \cdot)$ from the first $n_c = \gamma n$ samples of every trajectory, with $\gamma \in (0, 1)$. As in the proof of Proposition~\ref{prop:loss}, the best grid point under $\hat J_{n_c}$ loses at most $2e_{n_c}$ in $J_{h,L}$ against the best grid point. The coarse search keeps the $K$ best grid points, which makes the result robust to ranking errors between similar designs and to local maxima.

The local search refines each kept point at the full length $n$, as in pattern search~\citep{kolda2003optimization}. For a step $\Delta > 0$, it moves from a design $C$ to the best feasible point of $C + \{-\Delta, 0, \Delta\}^m$, and it stops when $C$ itself is that point. Every move strictly increases $\hat J_n(\theta; \cdot)$, and the reachable points form a finite grid, so the search stops after finitely many moves. The best refined design is $\hat C_{m,n}$. The tolerance $\varepsilon_{\rm opt}$ that it attains can be reduced by a finer grid, a larger $K$, and a smaller step $\Delta$. Any other method that attains~\eqref{eq:design-rule} can replace the two searches.

The two searches are repeated for the threshold candidates $m = 1, \dots, M$, and the selected number of thresholds $\hat m$ maximizes $\hat J_n(\theta; \hat C_{m,n})$ over $m \le M$. The information rate of the quantized output is at most that of the unquantized output $y_t$ for every $m$, whereas the cost $\log_2(m+1)$ of an observation grows with $m$. The criterion per bit~\eqref{eq:perbit} therefore tends to zero as $m$ grows, thus a small bound $M$ suffices. Algorithm~\ref{alg:design} summarizes the procedure.

\begin{algorithm}[t]
\caption{Quantizer design at a parameter value $\theta$}
\label{alg:design}
\begin{algorithmic}[1]
\Require parameter value $\theta {} \in \Theta_h$; step $h$ and directions $\xi_j$; depth $L$; trajectory length $n$; criterion $\Psi$; bound $M$; coarse grids $\mathcal G_m \subset \mathcal C_m$; prefix length $n_c < n$; number $K$ of kept designs; local step $\Delta$
\State Generate $Y, \tilde Y$ at $\theta$ and $Y_j^{\pm}$ at $\theta \pm h\xi_j$, $j = 1, \dots, r$
\For{$m = 1, \dots, M$}
    \State Keep the $K$ designs in $\mathcal G_m$ with the largest $\hat J_{n_c}(\theta; C)$
    \For{each kept design $C$}
        \Repeat
            \State $C \gets$ a maximizer of $\hat J_n(\theta; \cdot)$ over the feasible points of $C + \{-\Delta, 0, \Delta\}^m$, $C$ itself if it is one
        \Until{$C$ is unchanged}
    \EndFor
    \State $\hat C_{m,n} \gets$ the refined design with the largest $\hat J_n(\theta; C)$
\EndFor
\State \Return $\hat m \in \argmax_{m} \hat J_n(\theta; \hat C_{m,n})$ and $\hat C_{\hat m, n}$
\end{algorithmic}
\end{algorithm}

\subsection{Two-stage design}
\label{sec:twostage}

Problem~\ref{prob:design} is posed at the true value $\theta_0$, which is unknown, so Algorithm~\ref{alg:design} cannot be run at $\theta_0$ itself. The two-stage design~\citep{pronzato2013design} therefore records the data in two stages. The first stage records observations with default thresholds $C^{(0)}$, namely $m_0$ thresholds that make $\theta$ identifiable. An estimate $\hat\theta_1$ of $\theta_0$ is computed from these observations. Algorithm~\ref{alg:design} at $\theta = \hat\theta_1$ then selects the number of thresholds $\hat m$ and the thresholds $\hat C := \hat C_{\hat m, n}$ by simulation, without further plant data. The second stage records observations with $\hat C$. The final estimate of $\theta_0$ uses the data of both stages, for instance by maximum likelihood computed with expectation maximization~\citep{godoy2011identification, aguero2017em}. Because $\hat C$ depends only on the first-stage data, the likelihood of the data of both stages is computed as if both quantizers had been fixed in advance, as in adaptive input design~\citep{gerencser2009identification}.

A fraction $\alpha \in (0, 1)$ of the bit budget $R$ is allocated to the first stage, so the first and the second stage record $N_1 = \alpha R/\log_2(m_0+1)$ and $N_2 = (1-\alpha)R/\log_2(\hat m+1)$ observations, respectively. A larger $\alpha$ spends more bits with $C^{(0)}$, whereas a smaller $\alpha$ makes $\hat\theta_1$, and hence $\hat C$, less accurate. Because the accuracy of $\hat\theta_1$ depends on $N_1$, a larger budget allows a smaller $\alpha$. Repeating the design and the second stage with updated estimates gives a multistage design. Algorithm~\ref{alg:twostage} summarizes the two-stage design.

The designs $\hat C_{m,n}$ of Algorithm~\ref{alg:design} are consistent in an iterated limit, up to the tolerance $\varepsilon_{\rm opt}$, provided that the convergence in~\eqref{eq:limit} is uniform on $\mathcal C_m$. In this case, $e_n \to 0$ almost surely, and Proposition~\ref{prop:loss} implies that the loss of $\hat C_{m,n}$ is asymptotically at most $\varepsilon_{\rm opt}$ plus a bias term of order $h^2 + \rho^L$, which vanishes as $h \to 0$ and $L \to \infty$. For the two-stage design with $\Psi = \lambda_{\min}$, let $\hat\theta_1$ be consistent and $J$ be continuous in $\theta$ uniformly in $C$. In the limit above with $\varepsilon_{\rm opt} \to 0$, and as $R \to \infty$ with $\alpha \to 0$ and $\alpha R \to \infty$, the minimum eigenvalue of the information of both stages, divided by $R$, then tends to the maximum of $J(\theta_0; \cdot)$.

\begin{algorithm}[t]
\caption{Two-stage quantizer design}
\label{alg:twostage}
\begin{algorithmic}[1]
\Require bit budget $R$; fraction $\alpha \in (0, 1)$; default thresholds $C^{(0)}$, with $m_0$ thresholds, that make $\theta$ identifiable; estimator $\hat\theta(\cdot)$; the inputs of Algorithm~\ref{alg:design} except $\theta$
\State Record $N_1 = \lfloor \alpha R / \log_2(m_0 + 1) \rfloor$ observations $z^{(1)}$ with $C^{(0)}$, and let $\hat\theta_1$ be $\hat\theta(z^{(1)})$ projected onto $\Theta_h$
\State Run Algorithm~\ref{alg:design} at $\theta = \hat\theta_1$, which returns $(\hat m, \hat C)$
\State Record $N_2 = \lfloor (1 - \alpha) R / \log_2(\hat m + 1) \rfloor$ observations $z^{(2)}$ with $\hat C$
\State \Return $\hat\theta$ computed from $z^{(1)}$ and $z^{(2)}$ jointly, for instance by maximum likelihood
\end{algorithmic}
\end{algorithm}

\section{Numerical Study}
\label{sec:numerics}

Two systems are studied. First, on a quantized ARX system, for which reference values of the information can be computed, the designs of Algorithm~\ref{alg:design} are compared with the reference optimum for each $m$, and the convergence of the estimate is checked. Second, a model of cascaded tanks, for which only trajectories can be simulated, shows that the method also applies to a realistic system. On both systems, the selected designs are used to identify $\theta_0$, and their errors are compared with those of a threshold at the median of the output. This threshold is optimal for a location parameter in i.i.d.\ Gaussian noise~\citep{ribeiro2006bandwidth}. All designs use the criterion per bit~\eqref{eq:perbit} with $\Psi = \lambda_{\min}$ and at most $M = 3$ thresholds.

\subsection{Quantized ARX system}
\label{sec:num-arx}

\subsubsection{Model and reference values}
The first system has the parameter $\theta = (a, b)$ with $\lvert a \rvert < 1$ and is described by
\begin{equation}
y_t = a\, y_{t-1} + b\, u_t + w_t,
\label{eq:arx}
\end{equation}
where $u_t$ is i.i.d.\ and uniform on $\{\pm1\}$ and $w_t \sim N(0, \sigma_w^2)$. It is a special case of~\eqref{eq:system} with $x_t = y_t$ and $v_t = 0$. The true value is $\theta_0 = (a, 1)$ in six settings, namely $\sigma_w \in \{1, 0.3\}$ and $a \in \{0, 0.5, 0.9\}$. Because the hidden state $y_{t-1}$ is scalar, the likelihood of the observations can be computed by a point-mass filter, which represents the law of $y_{t-1}$ on a fine grid~\citep{bucy1971digital}. The reference Fisher information $I(\theta_0; C)$ is then the average of the outer products of the conditional scores of $q_t$ over many simulated observations, accurate to $0.5\%$ of the exact value. 
For $a = 0$, the system reduces to the one-tap FIR system of~\citet{wang2008space}, whose information has a closed form, and the reference values for $a = 0$ agree with it within $0.5\%$. The point-mass filter needs the transition density and a scalar state, so it serves only as a reference.

\subsubsection{Design by Algorithm~\ref{alg:design}}
Algorithm~\ref{alg:design} is run at $\theta_0$ with trajectories of length $n = 2^{18}$. The setting $a = 0.9$, $\sigma_w = 0.3$, in which the output varies slowly compared with the noise, is called the \emph{hard case}, and it uses $n = 2^{20}$. The depth $L$ is increased until the estimate saturates. Two modifications are made in practice. To reduce the variance, the estimates $v_j$ are averaged over 10 independent sets of the trajectories $Y, \tilde Y, Y_j^{\pm}$. To reduce the computation, the coarse search uses successive halving~\citep{karnin2013almost} instead of a single prefix length $n_c$, that is, it evaluates the grid on prefixes of increasing length and discards the worse half of the designs after each round. Averaging does not change the limit criterion $J_{h,L}$, and successive halving is only another search, so Proposition~\ref{prop:loss} applies unchanged. The grids and the local step were chosen in preliminary simulations that used no reference value.

\begin{figure}[t]
\centering
\includegraphics[width=\columnwidth]{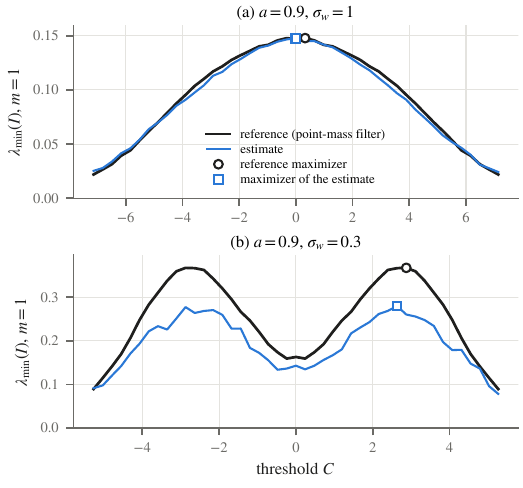}
\caption{Criterion for $m = 1$ on~\eqref{eq:arx} with $a = 0.9$, reference and estimated.}
\label{fig:arx}
\end{figure}

Because the criterion for $m = 1$ depends on a single threshold, it can be drawn as a curve and compared with the reference values directly. In the two settings with $a = 0.9$, namely with $\sigma_w = 1$ and in the hard case, the estimate from the trajectories above and the reference values are shown in Fig.~\ref{fig:arx}. For $\sigma_w = 1$, the estimate follows the reference curve. In the hard case, the estimate lies about $20\%$ below the reference curve, mainly because of the bias of the depth-$L$ limit. Its maximizer nevertheless lies on the same peak as the reference maximizer and loses $0.2\%$. The ranking of the designs is thus more accurate than the values, because by Proposition~\ref{prop:loss} the loss depends on the bias only through its difference between the selected and the best design.

\begin{figure}[t]
\centering
\includegraphics[width=\columnwidth]{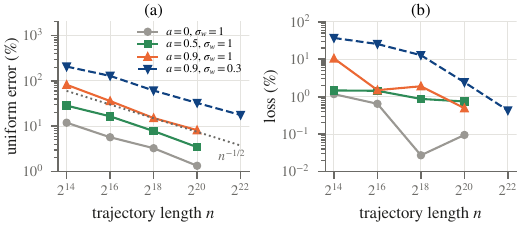}
\caption{Convergence of the estimate for $m = 1$ on~\eqref{eq:arx}. (a) Uniform error $e_n$ relative to the median of $J_{h,L}$. (b) Loss of the maximizer of the estimate. Means over 4--10 sets of trajectories.}
\label{fig:arx-conv}
\end{figure}

For increasing trajectory lengths $n$, the criterion for $m = 1$ is then estimated with the depths above in the four settings with $\sigma_w = 1$ or $a = 0.9$. It is compared with the limit criterion $J_{h,L}$ of~\eqref{eq:limit}, which the point-mass filter computes from the $L - 1$ past symbols of each context. The uniform error $e_n$ of Section~\ref{sec:search} and the loss of the maximizer of the estimate are shown in Fig.~\ref{fig:arx-conv}. In all four settings, the uniform error $e_n$ decreases about as $n^{-1/2}$, and the loss of the maximizer of the estimate falls below $1\%$ at the largest $n$. These results support the condition $e_n \to 0$ of Section~\ref{sec:twostage}.

\begin{table}[t]
\caption{Quantized ARX system~\eqref{eq:arx} at $\theta_0 = (a, 1)$. On the left, the reference criterion per bit $J^\star_m$ of the best $m$ thresholds, with the best $m$ in bold, and the gain of the best $m$ over $m = 1$. On the right, the design $\hat C$ selected by Algorithm~\ref{alg:design}, with $\hat m$ thresholds, its estimated and reference criterion, and its relative loss.}
\label{tab:arx}
\centering\footnotesize\setlength{\tabcolsep}{2.6pt}
\begin{tabular}{cc ccc r c ccr}
\hline
 & & \multicolumn{4}{c}{reference optimum} & \multicolumn{4}{c}{Algorithm~\ref{alg:design}} \\
$\sigma_w$ & $a$ & $J^\star_1$ & $J^\star_2$ & $J^\star_3$ & gain & $\hat m$ & $\hat J_n(\theta_0; \hat C)$ & $J(\theta_0; \hat C)$ & loss \\
\hline
1 & 0 & \textbf{0.441} & \textbf{0.440} & 0.405 & tie & 2 & 0.437 & 0.440 & $0.2\%$ \\
1 & 0.5 & 0.294 & \textbf{0.369} & 0.360 & $25\%$ & 2 & 0.366 & 0.368 & $0.2\%$ \\
1 & 0.9 & 0.149 & 0.197 & \textbf{0.214} & $44\%$ & 3 & 0.211 & 0.213 & $0.3\%$ \\
0.3 & 0 & 3.54 & \textbf{4.46} & 4.02 & $26\%$ & 2 & 4.42 & 4.42 & $0.9\%$ \\
0.3 & 0.5 & 1.08 & 1.77 & \textbf{2.18} & $101\%$ & 3 & 2.12 & 2.15 & $1.0\%$ \\
0.3 & 0.9 & 0.369 & 0.817 & \textbf{1.06} & $187\%$ & 3 & 1.08 & 1.04 & $2.0\%$ \\
\hline
\end{tabular}
\end{table}

The reference optimum for each $m$ and the design selected by Algorithm~\ref{alg:design} are given in Table~\ref{tab:arx}. Under the reference criterion, the best number of thresholds is two or three, except for a tie between one and two in one setting, and the gain of the best $m$ over $m = 1$ reaches $187\%$. The selected number of thresholds $\hat m$ is optimal in all six settings and for each of four random seeds. The selected designs $\hat C$ lose $0.2$--$2.0\%$ against the reference optimum $C^\star$ in the relative loss $1 - J(\theta_0; \hat C)/J(\theta_0; C^\star)$. Over the four seeds, the losses stay within $1\%$ except in the hard case, where the loss reaches $3.6\%$ because of the bias of the depth-$L$ limit.

\subsubsection{Identification}
The selected design is finally used to identify $\theta_0$ for $a = 0.9$, $\sigma_w = 1$, and compared with the median threshold $C = 0$. For each design and each bit budget $R \in \{2^{10}, \dots, 2^{14}\}$, 200 data sets of $R/\log_2(m+1)$ observations are simulated. From each data set, $\theta_0$ is estimated by maximum likelihood~\citep{aguero2017em}, with the likelihood computed by the point-mass filter and the iterative maximization initialized at the true value $\theta_0$. Because the mean squared error matrix $\mathrm{MSE}$ of $\hat\theta$ decreases about as $1/R$, the error is measured by $R\cdot\lambda_{\max}(\mathrm{MSE})$, averaged over the budgets, and the CRLB gives the lower bound $1/J(\theta_0; C)$ for it. The selected design reduces this error by about $30\%$ against the median threshold, as the bounds predict, and the errors of both designs lie within $10\%$ of their bounds.

When $\theta_0$ is unknown, the two-stage design of Algorithm~\ref{alg:twostage} with $\alpha = 1/4$ runs Algorithm~\ref{alg:design} at the first-stage estimate $\hat\theta_1$ with one set of trajectories. Measured in the same way, its error equals that of the design selected at $\theta_0$ within the Monte Carlo error, and it selects three thresholds in 978 of 1000 runs.

\subsection{Cascaded tanks}
\label{sec:num-tank}

\subsubsection{Model and reference values}
The second system is adapted from the cascaded-tanks benchmark of~\citet{schoukens2017three}, with level switches on the lower tank in place of the level sensor of the benchmark. Its state $x = (x_1, x_2)$ consists of the levels of the upper and lower tanks, and its input $u$ is the pump voltage. The system is described by
\begin{equation}
\begin{aligned}
\dot x_1 &= -0.5\sqrt{x_1} + 0.5\, u + 0.15\, w_1,\\
\dot x_2 &= 0.5\sqrt{x_1} - k_3\sqrt{x_2} + \sigma_2\, w_2,\\
y_t &= x_2(tT) + v_t,
\end{aligned}
\label{eq:tank}
\end{equation}
where $T = 6$ is the sampling period, $w = (w_1, w_2)$ is Gaussian white noise with $\E[w(s)\, w(s')^\top]$ equal to $\delta(s - s')$ times the identity matrix, and $v_t \sim N(0, 0.1^2)$ is i.i.d. Both $x_1$ and $x_2$ are reflected at $0$, and $x_2$ is capped at $12$. When $x_1$ exceeds $10$, the excess leaves the upper tank and half of it is added to $x_2$, so $x \in [0, 10] \times [0, 12]$. The input $u \in \{1, 2, 3\}$ is constant between the sampling instants $tT$ and follows a multilevel version of the generalized binary noise of~\citet{tulleken1990generalized}. At each sampling instant, it keeps its value with probability $0.6$ and is otherwise drawn uniformly from $\{1, 2, 3\}$. The input thus holds its value for several sampling periods, so that the levels approach their steady values. The $m$ switches are placed at the thresholds $C = (C_1, \dots, C_m)$ and report $q_t = \mathcal Q_C(y_t)$. The unknown parameter is $\theta = (\log k_3, \log\sigma_2)$ with $\theta_0 = (\log 0.5, \log 0.15)$.

Because of the square roots and the boundaries, the information $I(\theta; C)$ of~\eqref{eq:tank} has no closed form. The information is therefore estimated from trajectories of~\eqref{eq:tank}, which are simulated with an Euler discretization. A bootstrap particle filter~\citep{gordon1993novel} also needs no transition density, and it serves as the baseline. With this filter, the KL rates in~\eqref{eq:curvature} are computed from log-likelihoods instead of CTW models. The baseline is accurate with enough particles and data, so the two methods are compared in the CPU time needed to select a good design. The reference value of $I(\theta; C)$ is the value of the baseline with 1024 particles on long data sets.

\begin{table}[t]
\caption{Designs for the cascaded tanks~\eqref{eq:tank}. On the left, the criterion per bit, estimated by Algorithm~\ref{alg:design} and computed by the reference, with the largest in bold. On the right, the error $R\cdot\lambda_{\max}(\mathrm{MSE})$ of simulated maximum likelihood for the bit budgets $R \ge 2^{11}$, with $95\%$ bootstrap intervals, and its lower bound $1/J$ from the CRLB.}
\label{tab:tank}
\centering\footnotesize\setlength{\tabcolsep}{2.9pt}
\begin{tabular}{cl cc cc}
\hline
 & & \multicolumn{2}{c}{criterion per bit} & \multicolumn{2}{c}{$R\cdot\lambda_{\max}(\mathrm{MSE})$} \\
$m$ & $C$ & estimate & reference & sim.\ ML & bound \\
\hline
1 & 4.1 (median) & 0.0389 & 0.0419 & 36.8 [33.4, 40.9] & 23.9 \\
1 & 4.83 & 0.0472 & 0.0526 & 31.8 [27.9, 35.8] & 19.0 \\
2 & 3.56, 4.57 & 0.0623 & 0.0705 & 21.4 [19.2, 23.8] & 14.2 \\
3 & 3.30, 4.07, 4.83 & \textbf{0.0703} & \textbf{0.0792} & 18.7 [16.8, 20.7] & 12.6 \\
\hline
\end{tabular}
\end{table}

\begin{figure}[t]
\centering
\includegraphics[width=\columnwidth]{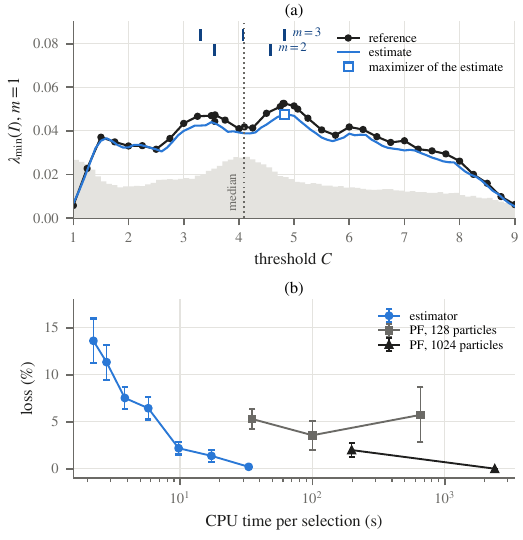}
\caption{Cascaded tanks~\eqref{eq:tank}. (a) Criterion for $m = 1$, computed by the reference and by the estimator. The shading is the stationary density of the output $y_t$, and the dotted line is its median. The ticks at the top mark the selected thresholds for $m = 2$ and $m = 3$. (b) Loss of the selected design against the single-core CPU time per selection, measured under the same conditions for the estimator and the particle filter (PF), for increasing amounts of data. Means and standard errors over 7--30 repetitions.}
\label{fig:tank}
\end{figure}

\subsubsection{Design by Algorithm~\ref{alg:design}}
Algorithm~\ref{alg:design} is run at $\theta_0$ with depth $L = 8$ and, as in Section~\ref{sec:num-arx}, with 10 sets of trajectories, each of $n = 2^{18}$ samples. For four random seeds, the selected thresholds $\hat C$ agree within $0.001$.

The criterion for $m = 1$ is shown in Fig.~\ref{fig:tank}(a). The estimate lies typically $7\%$ below the reference and has the same shape, with several peaks. The selected threshold $4.83$ lies at the maximum of the reference curve. The median threshold $4.1$ lies in a valley, and the selected threshold gains $26\%$ over it.

The thresholds selected for each $m$ and their criteria are given in the left part of Table~\ref{tab:tank}. Under the reference, the criterion per bit increases with $m$, so $m = 3$, the number selected by Algorithm~\ref{alg:design}, is the best among $m \le 3$. The selected design with $m = 3$ gains $51\%$ over that with $m = 1$ and $89\%$ over the median threshold.

\subsubsection{Comparison with the particle filter}
The estimator, with one set of trajectories, and the particle filter, with 128 or 1024 particles, select among the same 29 candidate designs. The particle filter uses common random numbers across parameter values and the finite-difference step that gives it the smallest loss. The loss is the relative loss of Section~\ref{sec:num-arx} under the reference, with $C^\star$ the best candidate under the reference. Each method is run with increasing amounts of data, and the loss of the selected design is shown against the CPU time in Fig.~\ref{fig:tank}(b). At a similar CPU time, the estimator loses $0.2\%$, whereas the particle filter with $128$ particles loses about $5\%$. With $128$ particles, the estimate of the particle filter has an error from the finite number of particles. This error does not decrease with the number of observations and differs between designs, so more data does not reduce the loss. With $1024$ particles, the particle filter selects the best candidate, but it needs about $70$ times the CPU time of the estimator. The particle filter must run a separate filter for every candidate and every parameter value, because its weights depend on the symbols of the candidate and its particles on the parameter value. The estimator instead counts the symbols of each candidate in trajectories simulated once, and its CPU time goes mainly to counting. The difference in CPU time therefore comes from replacing filtering by counting.

\subsubsection{Identification}
The designs selected for each $m$ and the median threshold are used to identify $\theta_0$, with 200 data sets for each bit budget. Because the likelihood has no closed form, $\theta_0$ is estimated by simulated maximum likelihood, which maximizes the particle-filter log-likelihood~\citep{kantas2015particle} by iterations initialized at $\theta_0$. The errors and their bounds are given in the right part of Table~\ref{tab:tank}. Against the median threshold, the design with $m = 1$ reduces the error $R\cdot\lambda_{\max}(\mathrm{MSE})$ by about $14\%$, but its $95\%$ interval overlaps that of the median threshold. The designs with $m = 2$ and $m = 3$ reduce the error by about $40\%$ and $50\%$, respectively, as the bounds predict, and their intervals do not overlap that of the median threshold. For all designs, the error is $1.5$--$1.7$ times its bound, with no trend in the bit budget $R$. Simulated maximum likelihood is not efficient here, because the number of particles stays fixed while the variance of the particle-filter log-likelihood grows with the number of observations~\citep{kantas2015particle}. This loss of efficiency is similar for all designs, so the reductions above follow those of the bounds.

\section{Conclusion}
\label{sec:conclusion}
This paper turns quantizer design for systems with memory into a problem of numerical estimation of the FIM, and it solves this problem with an estimator that needs only simulated trajectories. The two examples show effects that the closed-form results do not show. With fixed thresholds, the number of thresholds that is best per bit often exceeds one. The common threshold at the median of the output can lie in a valley of the criterion. The examples also show that the ranking of the designs remains accurate when the estimated criterion is biased, in agreement with the loss bound. The consistency of the design is currently conditional on the uniform convergence of the estimate. The proof of this condition, the rate of convergence of the design, and the choice of the step size and of the depth are left for future work.

\section*{DECLARATION OF GENERATIVE AI AND AI-ASSISTED TECHNOLOGIES IN THE WRITING PROCESS}
During the preparation of this work the authors used Claude (Anthropic) in order to improve the readability and language of the manuscript. After using this tool, the authors reviewed and edited the content as needed and take full responsibility for the content of the publication.

\bibliography{reference}

\end{document}